# Machine learning and AI research for Patient Benefit: 20 Critical Questions on Transparency, Replicability, Ethics and Effectiveness

**Sebastian Vollmer**[1,2]*
PhD

**Bilal A. Mateen**[1,3,4]*
MBBS

**Gergo Bohner**[1,2]
B.S.

**Franz J Király**[1,5]
PhD

**Rayid Ghani**[6]
PhD

**Pall Jonsson**[7]
PhD

**Sarah Cumbers**[8]
PhD

**Adrian Jonas**[9]
PhD

**Katherine S.L. McAllister**[9]
PhD

**Puja Myles**[10]
PhD

**David Granger**[11]
MSc

**Mark Birse**[11]
MSc

**Richard Branson**[11]
MSc

**Karel GM Moons**[12]
PhD

**Gary S Collins**[13]
PhD

**John P.A. Ioannidis**[14]
DSc

**Chris Holmes**[1,15]
PhD

**Harry Hemingway**[16,17,18]
MD

* These authors contributed equally to the manuscript.
**Affiliations**

1) The Alan Turing Institute, Kings Cross, London NW1 2DB, United Kingdom
2) Departments of Mathematics and Statistics University of Warwick, Coventry CV4 7AL, United Kingdom
3) Warwick Medical School, University of Warwick, Coventry CV4 7AL, United Kingdom
4) Kings College Hospital, Denmark Hill, London, SE5 9RS, United Kingdom
5) Department of Statistical Science, University College London, Gower Street, London WC1E 6BT, United Kingdom
6) University of Chicago, 1155 E. 60th Street, Chicago, IL 60637
7) Science Policy and Research, National Institute for Health and Care Excellence. Level 1A, City Tower, Piccadilly Plaza, Manchester, M1 4BT, UK
8) Health and Social Care Directorate, National Institute for Health and Care Excellence. 10 Spring Gardens, London, SW1A 2BU
9) Data and Analytics Group, National Institute for Health and Care Excellence. 10 Spring Gardens, London SW1A 2BU, United Kingdom
10) Clinical Practice Research Datalink, Medicines and Healthcare products Regulatory Agency, Canary Wharf, London E14 4PU, United Kingdom
11) Medicines and Healthcare products Regulatory Agency, Canary Wharf, London E14 4PU, United Kingdom
12) Julius Centre for Health Sciences and Primary Care, UMC Utrecht, Utrecht University, Utrecht, The Netherlands
13) UK EQUATOR Centre, Centre for Statistics in Medicine, NDORMS, University of Oxford, Oxford, OX3 7LD.
14) Meta-Research Innovation Centre at Stanford (METRICS), Stanford University, Stanford, CA, USA
15) Department of Statistics, University of Oxford, Oxford, UK
16) Health Data Research UK London, University College London, 222 Euston Road, London NW1 2DA, UK
17) Institute of Health Informatics, University College London, 222 Euston Road, London NW1 2DA
18) The National Institute for Health Research, University College London Hospitals Biomedical Research Centre, University College London, 222 Euston Road, London NW1 2DA, UK

**Corresponding Author**
Prof. Chris Holmes
Address: Department of Statistics, University of Oxford, Oxford, UK
Email: cholmes@stats.ox.ac.uk
Telephone: 01865285874

# Abstract

Machine learning (ML), artificial intelligence (AI) and other modern statistical methods are providing new opportunities to operationalize previously untapped and rapidly growing sources of data for patient benefit. Whilst there is a lot of promising research currently being undertaken, the literature as a whole lacks: transparency; clear reporting to facilitate replicability; exploration for potential ethical concerns; and, clear demonstrations of effectiveness. There are many reasons for why these issues exist, but one of the most important that we provide a preliminary solution for here is the current lack of ML/AI-specific best practice guidance. Although there is no consensus on what best practice looks in this field, we believe that interdisciplinary groups pursuing research and impact projects in the ML/AI for health domain would benefit from answering a series of questions based on the important issues that exist when undertaking work of this nature. Here we present 20 questions that span the entire project life cycle, from inception, data analysis, and model evaluation, to implementation, as a means to facilitate project planning and post-hoc (structured) independent evaluation. By beginning to answer these questions in different settings, we can start to understand what constitutes a good answer, and we expect that the resulting discussion will be central to developing an international consensus framework for transparent, replicable, ethical and effective research in artificial intelligence (AI-TREE) for health.

# Introduction

Machine learning (ML), artificial intelligence (AI) and other modern statistical methods are providing new opportunities to operationalize previously untapped and rapidly growing sources of data for patient benefit. The potential uses include improving diagnostic accuracy [1], more reliably predicting prognosis [2], targeting treatments [3], and increasing the operational efficiency of health systems [4]. Examples of potentially disruptive technology with early promise include image-based diagnostic applications of ML/AI, which have shown the most early clinical promise (e.g. deep learning-based algorithms significantly improving accuracy in diagnosing retinal pathology compared to that of specialist physicians [5]), or natural language processing used as a tool to extract information from structured and unstructured (i.e. free) text embedded in electronic health records (e.g. patients' notes) which has also shown great potential [2]. Although, we are only just beginning to understand the wealth of opportunities afforded by these methods, there is growing concern in the academic community that because the products of these methods are not perceived in the same way as other medical interventions (e.g. pharmacological therapies), they do not have well-defined guidelines for development and use and rarely undergo the same degree of scrutiny.

*The Need for Guidance Today*

Increasingly clinicians are involved in the process of generating, using and interpreting clinical applications of machine learning and artificial intelligence. This has highlighted a series of fundamental issues that need to be addressed for clinicians and policy-makers to be able to make informed decisions about which tools are likely to be safe and effective. While many best practice recommendations for design, conduct, analysis, reporting, impact assessment and clinical implementation can be borrowed from the traditional biostatistics and medical statistics literature (e.g. [6]), they are not sufficient to guide the use of ML/AI research. Several high profile publications have demonstrated that there is a lack of transparency [7,8], replicability [9], ethics [10] & effectiveness [11] (TREE) in the reporting & assessment of ML/AI-based prediction models.

Furthermore, nearly all biomedical research publications using ML/AI are reported in a 'positive' light, however, there are currently few examples, if any, of health-related ML/AI algorithms that have conclusively demonstrated improved patient outcomes. So how do we (i.e. clinicians, methodologists, statisticians, data scientists, and healthcare policy maker), identify and critique relevant ML/AI research to assess its suitability for translation from the 'computer-bench' to the bedside? Currently, such assessments are difficult because we lack guidance on what differentiates good from bad at the different stages of the healthcare-related ML/AI development pipeline, from design and data analysis, through to reporting, and evaluation of effectiveness/impact on health. Producing such guidance is a major undertaking due to the ever-growing battery of ML/AI algorithms and the multifaceted nature of assessing performance and clinical impact. Not taking action is unacceptable, and if we wait for a more definitive solution, we risk wasting valuable work [12-16], whilst allowing futile research to continue unchecked, or worse, translation of ineffective (or even harmful) algorithms into clinical practice.

*An Initial Framework - 20 Critical Questions for Health-related ML/AI Technology (Box 1)*

Ensuring any potential guidance is applicable to all relevant stakeholders in this emerging field requires representatives of each part of the health data-science cycle (Figure 1) to be involved in its creation. Here we propose a series of critical questions to help identify common pitfalls which can undermine ML/AI-based applications in health. These questions arose as a result of a dialogue between a group of UK-based and international organisations spanning a range of different perspectives, from statutory regulators and national advisory bodies, to academic organisations, including:

- The UK's National Data Science Institute (The Alan Turing Institute [17]), established 2015;
- The UK's Health Data Science at Scale Initiative (HDR UK [18]), established 2018;
- The UK's National Institute for Clinical and Care Excellence (NICE [19]); established in 1999;
- The UK's Medicines and Healthcare products Regulatory Agency (MHRA [20])), established in 2003 (as a merger of the Medicines Control Agency (MCA) and the Medical Devices Agency (MDA));
- The UK's Clinical Practice Research Datalink (CPRD [21]), supported by the Medicines and Healthcare products Regulatory Agency and National Institute of Health Research, is a UK Government research service providing anonymised UK primary care data research since the late 1980s.
- The Enhancing the Quality and Transparency of Health Research (EQUATOR) Network [22], established in 2006;
- The Meta-Research Innovation Centre at Stanford (METRICS) [23], established in 2014;
- The Data Science for Social Good (DSSG) Program [24], established in 2013 and run by the Centre for Data Science and Public Policy at the University of Chicago

Initial discussions regarding the principles on which standards for ML research might be based identified a series of themes central to high quality health-related research and technologies: transparency, replicability, ethics & effectiveness (TREE). Based on these themes, we have developed an initial framework - 20 critical questions (Box 1) which can be used as a guide. The questions are not only relevant for those who use the findings (i.e. patients and policy makers), but also those who generate ML/AI health research. For the former group these questions might inform development of critical appraisal and reporting guidelines. For the latter, these questions might inform the way that research groups design and conduct such research. We hope that adoption of this framework, and other related publication which are in the pipeline e.g. NHS-England's AI Code of Conduct [25], will help to build trust in the underlying processes and results of health-related ML/AI research.

## *Figure 1: The Data Science Cycle for Healthcare*

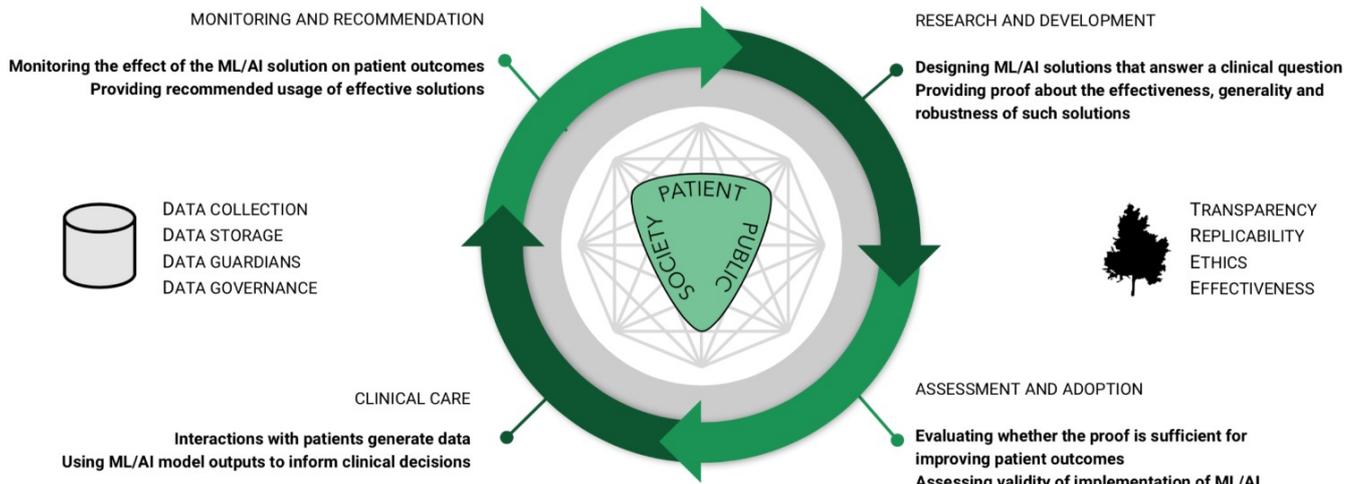

*Figure 1 Legend:* This figure summarises the stages relevant for and affected by ML/AI methods and their adoption in clinical care. The collaborative effort behind this piece represents all stages required for developing transparent, reproducible, ethical and effective ML/AI solutions for clinical care. For almost all projects this process will not be linear from start to finish, rather, requiring a continuous feedback loop between the different stakeholders responsible for the various tasks. Starting with the vital input for any algorithmic solution, data, this can be generated either via planned experiments or collected during routine clinical practice. Given the data, the necessary data guardians then need to assess any proposed projects, provide the necessary access and oversee the users' activities in order to ensure the safe and efficient use of data. The research and development phase puts an idea into practice, and provides examinable proof that an algorithm indeed accomplishes the stated goals. Whether a solution is truly fit for purpose (i.e. transparent, reproducible, ethical and effective) should be first assessed by independent bodies. Finally, the performance of the ML/AI solutions should be continually monitored to demonstrate that it remains effective in improving patient outcomes, after it becomes recommended standard practice. The implementation of any new algorithm will influence the data collected at the level of care, by assigning increased importance in those features that are used as model inputs, whilst generating new information (i.e. the model outcomes) in routine practice, thus completing the development cycle, and initiating a new cycle.

## Box 1: 20 Critical Questions for Health-related ML/AI Technology

*Overall Project-specific Question*

1) How is the ML/AI model embedded in feedback loops as part of a learning health system?

*Inception-specific Questions*
2) What is the health question relating to patient benefit?
3) When and how should patients be involved in data collection, analysis, deployment, and use?
4) Is there organisational transparency about the flow of data?

*Analysis-specific Questions*
5) Is the data suitable to answer the clinical question, i.e. does it capture the relevant real-world heterogeneity, and is it of sufficient detail and quality?
6) Does the methodology reflect the real-world constraints and operational procedures associated with data collection and storage?
7) On what basis are data accessible to other researchers?
8) What computational and software resources are available, and are they sufficient to tackle this problem?
9) Are the reported performance metrics relevant for the clinical context in which the model will be used?
10) Is the reported gain in statistical performance with the ML/AI algorithm clinically justified in the context of any trade-offs?
11) Is the ML/AI algorithm compared to the current best technology, and against other appropriate baselines?
12) Are the different parts of the prediction modelling pipeline available to others to allow for methods reproducibility, including: the statistical code for 'pre-processing', and the modelling workflow (including the methods, parameters, random seeds, etc. utilised)?
13) Are the results reproducible in settings beyond where the system was developed (i.e. external validity)?

*Impact Evaluation-specific Questions*
14) What evidence is there that the model does not create or exacerbate inequities in healthcare by age, sex, ethnicity or other protected characteristics?
15) What evidence is there that clinicians and patients find the model and its output (reasonably) interpretable?
16) What evidence is there of real world model effectiveness in the proposed clinical setting?

*Implementation-specific Questions*
17) Is the model being regularly re-assessed, and updated as data quality and clinical practice changes (i.e. post-deployment monitoring)?
18) Is the ML/AI model cost-effective to build, implement, and maintain?
19) How will the potential financial benefits be distributed if the ML/AI model is commercialized?
20) How have the regulatory requirements for accreditation/approval been addressed?

# 20 Critical Questions

*Overall Project-specific Question*

**1. How is the ML/AI model embedded in feedback loops to facilitate a learning health system?**

'Lone wolf' ML/AI may occasionally promote innovation; however, it is being increasingly recognised that such research needs to be seen in a wider organizational context to be made most useful. This context is best understood as a learning health system in which data science is used to inform clinical decision making (and health system operations) in an iterative feedback loop; a move away from the classical framework of a linear pathway with an 'end user'.

At the local level, this involves a range of steps to achieve clinical adoption, including: integration into current IT systems, human resources, financial investment, etc. At the societal level, the ML/AI 'revolution' in health has resulted in major organizational change, including new national research organisations and initiatives in the UK (e.g. Alan Turing Institute [17], & Health Data Research UK [18]), new national initiatives (e.g. NHS workforce Topol review [26]), updated health technology assessment regulation (e.g. [27,28]), readiness preparation of guidance developers (e.g. NICE [19]) and the growth of a new (global) industrial sector (e.g. Google DeepMind, Benevolent AI, Babylon), as well as partnerships between health systems (e.g. The European Health Data Network [29]). Understanding to what extent these different organisations are engaged in a given piece of ML/AI research may help elucidate the pathways to impact in clinical practice. In summary, researchers should be cognisant of the path from development to implementation, as well as be able to describe which parts of the healthcare data science cycle their proposed research engages with. This doesn't preclude proof of concept work that may only occupy a small angle of the healthcare data science cycle, but equally this should be stated up front.

*Inception-specific Questions*

**2. What is the health question relating to patient benefit?**

The vast majority of published clinical prediction models are never used in clinical practice [30]. One reason for this is the lack of a specific clinical decision making process which the model may meaningfully inform or optimise; simply predicting future events on their own does not help a clinician do anything differently. In other words; just because we can, doesn't mean we should. Clarity is needed at the outset of research on the potential use of a proposed model (i.e., what the consequent actions would be, based on the model outcome and what patient outcomes are they improving), whether any underlying trade-offs that arise from the use of the model are in fact justified by the additional benefits provided by it and the likelihood of successful deployment in clinical practice.

**3. When and how should patients be involved in data collection, analysis, deployment, and use?**

With the growing use of routinely collected individual participant data (in addition to researcher collected data) often with an alternative legal basis (i.e. legitimate interests) to individual consent, it is more important now than ever, that patient and public involvement (PPI) is at the centre of the health data science cycle (figure 1). The responsibility of researchers to involve individuals whose data is being utilised in decisions about the use of said data has been highlighted by new legislation in Europe (GDPR), as well as the national standards for public involvement in research [31]. PPI can be at multiple stages in data science cycle, including but not limited to: identifying the need for an AI/ML algorithm, supporting the development of the algorithm, and determining the acceptability of the algorithm in practice. The exemption from seeking individual consent, does not mean that the researchers are exempt from engaging patients and public altogether, and thus, at the heart of every healthcare ML/AI project should be a clear understanding of the acceptability of the proposed model and outcomes to those from whom the data was collected, the users (i.e. clinicians) and the impacted individuals (i.e. those for whom the model will be used to inform clinical management).

**4. Is there organisational transparency about the flow of data?**

Patients have strong views about transparency in the flow of data, and how their data is secured [32]. For patients and their clinicians to trust ML/AI models it is important that they understand the interactions which led to the development of the model, whether they are between organisations in the public, not for profit and industrial sectors, or within them (e.g. transfer from one hospital department to another). Complying with the aforementioned legislative frameworks (e.g. GDPR) whilst necessary, is not sufficient to demonstrate the transparency required to produce trustworthy ML/AI research. The degree of detail necessary will differ depending on the institutions involved, and the nature of the work being undertaken. Therefore, the responsibility lies with ML/AI algorithm developers and those involved in accessing, transferring, or storing the data, to engage key stakeholders to understand what is required in each particular case. One aspect of the reporting procedure that can help ensure transparency regarding the aforementioned interactions is the inclusion of clear declarations of interest by all involved parties.

## *Analysis-specific Questions*

**5. Is the data suitable to answer the clinical question, i.e. does it capture the relevant real-world heterogeneity, and is it of sufficient detail and quality?**

The key question here is whether it is possible to answer the clinical question with the data available. For example, a dataset not containing the (known) relevant/important predictors of an outcome, is unlikely to satisfactorily answer questions about it. No ML/AI algorithm can produce something from nothing - this is worth pointing out as often there is an implicit belief to the contrary where it specifically concerns ML and AI algorithms. Data accuracy, sampling of participants, eligibility criteria, and missing data, also need to be considered when assessing the potential of developing useful and generalizable ML/AI algorithms. If data are available, but they are of poor quality or not relevant, it is unlikely that a good ML/AI application can be developed.

One should also consider the likelihood of failure when operating outside the training data range, e.g., an image-recognition/self-driving-car-decision-making system may fail when encountering for the first time a cyclist at night. Hence it is important to check whether the data- including timescale, heterogeneity (differences in data collection e.g. measuring devices, compliance), population, and situation - is in accordance with, and representative of, the envisioned clinical application scenario. These are just some of the issues that need to be addressed in determining whether the data are of sufficient quality and detail to be able to inform the clinical question of interest. However, every point around the health data science cycle (Figure 1) can inform the data collection mechanism and thus can be improved iteratively.

**6. Does the validation methodology reflect the real-world constraints and operational procedures associated with data collection and storage?**

Increasingly ML/AI research is making use of routinely collected data, including: data that exists as part of the health system (including electronic health records, clinical imaging, and genomic information); civil administrative data (e.g. death records, and educational achievement); and data from mobile and wearable devices (e.g. [33]). Information from these sources can arrive in batches, or via a continuous stream, and is often stored in different locations requiring reconciliation and collation, which in and of itself introduces a delay in when specific pieces of data are available for use. In contrast to these real-world constraints, ML/AI algorithms are often validated on historical data, yielding performance guarantees only under the assumption that the data generating process does not change, e.g., over time, or across hospitals. In practice, these assumptions are often severely violated and result in ML/AI models underperforming in deployment when compared to performance reported during system development [34]. A robust validation methodology in medicine and healthcare (and in most application areas of ML/AI) usually needs to take time into account and create temporally disjoint training and test sets [35,36] which account for how the data is collected and stored.

### 7. On what basis are data accessible to other researchers?

Data sharing is not an endpoint in itself but rather a means to enhance, verify and distribute the knowledge generated by the ML/AI algorithm [37] and thus is closely related to reproducibility and replicability. Most major funding sources now require applicants to outline a data management and data sharing plan; this can entail among other things, storing the data in a convenient format along with a data dictionary, a long-term archiving plan, and providing an independent access mechanism (e.g. a university ethics committee, or a R&D department). Where data used to develop the ML/AI algorithm have been accessed via national data custodians (e.g. CPRD [21]. NHS Digital [38], HQIP [39]), there are clear data access processes in place for independent validation by other researchers. Additionally, data sharing can be undertaken using a wide range of mechanisms, including:

1. Making the data available in open repositories such as datadryad.org [40] (after being anonymised using tools such as Amnesia [41]), or restricted access repositories such as UK data Archive [42];
2. Signing data sharing agreements;
3. Providing remote access to local computing facilities where the data is stored, as is possible with specific restricted access data enclaves such as NORC at the University of Chicago [43], and the electronic Data Research and Innovation Service (eDRIS) [44].

The advent of these facilities means that there are fewer reasons to be unable to share data from publicly funded research with other researchers. As such, ML/AI algorithm development should be accompanied by clear statements of what tools and mechanism will be used to support access to the data utilised, for the purposes of replication and validation of reported results.

### 8. What computational and software resources are available, and are they sufficient to tackle this problem?

There are many areas of health-related prediction modelling where it is common to work with millions of parameters, such as image-based deep learning [45], and statistical genetics [46]. As such, it has become common practice to not only determine the complexity of the data one is working with, but also the computational resources available, because much more often than with traditional statistical models, the latter can be the limiting factor in determining what analyses can be undertaken [47]. Simply put, in some situations more computational resources may in fact allow one to train a better model. For example, without compute servers with GPUs it is infeasible to even attempt to use complex neural network-based models, especially since these large-scale models require additional complex operations (e.g. regularisation) to prevent overfitting [48,49]. Similar issues can arise when using secure compute environments such as data enclaves or data safe havens, where the relevant software frameworks might not be available and thus would warrant implementation from scratch, or where there are limits to the computational resources available. Ideally, analysis would not be limited by the availability of computational resources, but developers need to understand the constraints within which they are working so that any analysis can be tailored to requirements. Furthermore, it is also important to understand the implications of using specific software, as the underlying licence can have far reaching consequences on the commercial potential and other aspects of the algorithms future (see [50] for a brief overview of software licencing for scientist-programmers).

## 9. Are the reported performance metrics relevant for the clinical context in which the model will be used?

The choice of metric matters in order to translate good performance in the (training data) evaluation setup to good performance in the eventual clinical setting with patient benefit. This discrepancy in model performance may arise for multiple reasons, but the most common of which being that the evaluation metrics are not good proxies for demonstrating improved outcomes for patients (e.g. misclassification error for a screening application with imbalanced classes). Another common mistake is choosing a performance metric which is vaguely related to, but not indicative or demonstrative of improved clinical outcomes for patients. For example, IBM's Watson expert system (Watson For Oncology (WFO) [51]), is used in several hospitals around the world to support decision making. However, published works describing WFO do not report relevant statistical (e.g. discrimination, calibration, etc.), and clinically-oriented (e.g. net benefit type) performance metrics. Instead they focus on concordance (true positive rate where the ground truth is provided by physician, i.e. the proportion of cases where WFO's recommendation agrees with that of the treating physician [e.g. 52-54]). Determining which is the appropriate performance metric to report will depend not only on the statistical task, but also the clinical decision making process to which the results will be applied, as such, it is important that all relevant parties (e.g. patients, data scientists, clinicians, etc.) are consulted in the selection process.

## 10. Is the reported gain in statistical performance with the ML/AI algorithm clinically justified in the context of any trade-offs?

For a new diagnostic or prognostic tool to be justified for routine use, it must offer a clinically meaningful advantage over existing approaches in addressing a specific need [55], which requires the use of an appropriate performance metric as discussed previously. Whilst necessary, the presence of a clinically meaningful advantage alone is not sufficient justification, as any improvement must be weighed against the cost of any changes it necessitates (e.g. the resource requirement to collect additional data). A recent paper published by Google investigating the accuracy of deep learning methods in combination with electronic health records for predicting mortality, readmission and length of stay is an apt demonstration of why this point is so important [2]. In their appendix, the authors provide a comparison of their deep learning (DL) model against a logistic regression model. The AUC improvement reported for each of the three tasks ranged from 0.01 to 0.02. If we assume that all caveats pertaining to statistical significance, and the sufficiency of the reported metric for making this next decision are met, the question that must therefore be asked is - whether the marginal gain of implementing a complex AI/ML solution is worth it, and is the need any more effectively addressed by the deep learning model? Whilst the answer to that question will certainly be situation specific, any proposed use case will (at minimum) need to justify the following: 1) the cost of developing, deploying, utilising, and maintaining a deep learning model such as the one described relative to the improvement observed; and 2) the need for additional subsidiary models to clawback inferential capabilities lost in the transition away from a model with human interpretable coefficients (white-box).

**11. Is the ML/AI algorithm compared to the current best technology, and against other appropriate baselines?**

ML/AI algorithms can be viewed as health technologies, and at the design stage consideration should be given to the currently used approach that the algorithm may replace. For almost all clinical questions, there will be a well-accepted "standard approach" from decades of biostatistics research, e.g. proportional hazards models for survival modelling. For many clinical outcomes and settings, there will be existing, published models. Similarly, there will be model proxies for uninformed guessing, such as predicting the majority class in a classification task. One common way to exaggerate the benefit of AI/ML approaches is to avoid any comparison of AI/ML against null models or the currently used approach, and instead compare to sub-par competitors (including in appropriately or weakly developed statistical models), or to avoid a comparison altogether. This "weak comparator"/"strawman" bias has been generally demonstrated in reports of (traditional) new vs existing prognostic models [49]. One such example comes from a systematic review of proposed modifications to the Framingham risk score for predicting the risk of a heart attack within 10 years; the review found that most proposed alternatives had flaws in their design, analyses, and reporting that cast doubt on the reliability of the claims for improved prediction [56]. The parallels to pseudo-scientific argumentation vs evidence-based medicine are especially noteworthy, as they make potential flaws relatively easy to spot to a medical practitioner, e.g., it is instructive for this purpose to replace "treatment" with "algorithm", and "best practice" with "gold standard".

**12. Are the different parts of the prediction modelling pipeline available to others to allow for methods reproducibility [57], including: the statistical code for 'pre-processing', and the modelling workflow (including the methods, parameters, random seeds, etc. utilised)?**

Reproducibility of research has become a growing concern across many scientific domains [58,59], and in the ML/AI field, access to the underlying code and raw data is central to preventing and mitigating reproducibility concerns. A recent example of how concerns regarding reproducibility in medical modelling research have manifested comes from a review of studies published using the MIT critical care database, MIMIC, which illustrates the degree to which inadequate reporting can impact replication in the prediction modelling [9]. Specifically, the reproducibility issues that have been identified in the literature occur not only in attempts to re-create reported findings, but also in how authors report data characteristics, such as the inclusion and exclusion criteria used to arrive at the final population of interest. In the review, 28 studies based on the same core data set (MIMIC) predicting mortality were investigated, and two important results were identified: For more than half of the studies examined the reproduced sample size differed by more than 25% from the reported sample size due to insufficiently clear descriptions of the inclusion/exclusion criteria. The result of inadequate reporting was that in the replication, the use of off-the-shelf logistic regression and boosted trees on the reproduced samples, produced better results in 64% and 82% of the 28 studies respectively, than the ML/AI model reported in the original study. These issues could have been easily avoided by providing the project code, specifically that relating to data cleaning and pre-processing. It should be noted that the RECORD reporting guidelines for studies using routinely collected health data already recommend providing detailed information to this effect [60], and there are several potential solutions that exist to facilitate this process, including code sharing and project curation platforms such GitHub.

**13. Are the results reproducible in settings beyond where the system was developed (i.e. external validity [61])?**

It has become increasingly evident that there is a relative lack of validation studies for diagnostic and prognostic tools, which precedes the emergence of the ML/AI era [62]. Even when external validation studies are carried out, reductions in the predictive accuracy of models, relative to their original (development study) performance is expected [34,63]. Systematic reviews have repeatedly observed this latter phenomenon in the applications of classical statistical models to a variety of healthcare-related prediction tasks, from mortality-risk prediction in acute kidney injury patients [64], to falls risk prediction in the elderly [65]. It is unclear to what degree this phenomenon associated with *results reproducibility* (i.e. "replication," that is, the production of corroborating results in a new study, having followed the same experimental methods [66]), is a consequence of the inadequate reporting that has been observed in the modelling literature [61], or other related issues. Given the additional complexities introduced by ML/AI algorithms, it is important that developers play a proactive role in ensuring that sufficient information is provided to allow their models to undergo rigorous, but fair [67], external validation (ideally by independent investigators).

## *Impact Evaluation-specific Questions*

**14. What evidence is there that the model does not create or exacerbate inequities in healthcare by age, sex, ethnicity or other protected characteristics?**

Systematic testing for bias and fairness is the first step in making informed model selection decisions to minimize inequities that could be caused by the use of ML/AI algorithms [68]. One way in which ML/AI algorithms result in bias is by making disproportionate errors in different populations. Depending on how the AI/ML algorithm has been developed (including whether key populations, defined by sex, age, ethnicity are sufficiently represented in the data, and included in the training of the algorithm), can influence the predictive accuracy of the algorithm in different subgroups, and thus when these predictions are used to take actions on individuals, they can create or exacerbate inequities [69]. The issue of data that is not truly representative of the entire target population is particularly important [70], as it highlights the importance of fairness considerations at every point in the data science cycle (figure 1), from data collection/research to implementation and evaluation. Other examples of how these issues can manifest in the real world can be found in ProPublica's analysis of the COMPAS recidivism prediction tool [10] and the USAs diabetes screening criterion [71], which both illustrate variation in performance of an algorithm based on race.

It will often be the case that many of the ML/AI algorithms developed will indeed have bias, but it is important to compare that to the bias in the existing systems being used. The types of performance variation that need to be investigated for depend on the consequent actions (or 'interventions') that the algorithm is helping to decide between. If the interventions are very expensive or have unwanted side-effects, then we would want to minimize disparities in the number of false positive predictions from different subgroups, to prevent unnecessary harm. If the interventions are predominantly assistive, we should be more concerned with disparities in false negatives, to prevent individuals missing out on a potentially beneficial input. It is worth noting that the explanation above presupposes that a decision threshold has been set, which may sometimes be outside of a developer's remit. However, this does not exempt them from needing to demonstrate that when using 'sensible' thresholds, that the algorithm does not create or exacerbate inequalities. In fact, there have been several methodological developments in the area of fairness evaluation to support this type of analysis [e.g. 72.73], and it is important that machine learning/AI developers and health practitioners engage with these tools.

## 15. What evidence is there that clinicians and patients find the model and its output (reasonably) interpretable?

Clinical adoption of an algorithm depends on two main factors, its usefulness (i.e. clinical utility), and its trustworthiness. When the outputs of a prediction model do not directly answer a specific clinical question it ultimately limits its usefulness, as discussed in earlier questions; whereas models whose processing pipeline is difficult to explain and justify to a general audience will invariably limit the trust placed in its outputs [74], despite robust and demonstrated statistical gains. However, trust is not the only reason that interpretability in important [75]. Recent changes in legislation (e.g. GDPR) have introduced additional protections for individuals, including a right to an explanation for how a decision was made, where it pertains them [76], thereby creating a legal requirement to provide insight into the underlying decision-making process an algorithm learns.

We alluded to the interpretability-performance trade-off in question 11, as it pertains to weighing up the loss of interpretability in transitioning from a 'white-box' to 'black-box' model, in exchange for improved performance. However, as the authors of the deep learning for EHR study recognised and attempted [2], there are partial solutions, including both model-specific and model agnostic methods, which can be used to claw back interpretability when using ML/AI methods [77]. Given that legal and moral burden of explanation to establish trust will vary with the nature of the decision, i.e. applications of ML/AI in health that influence the allocation of potentially life-prolonging treatments will necessitate a much higher explanatory burden to satisfy those individuals who are affected, the sufficiency of any explanations and adequacy of any insight producing method can only be determined through consultation and collaboration with the end users (clinicians), and target audience (patients).

## 16. What evidence is there of real world model effectiveness in the proposed clinical setting and how are unintended consequences prevented?

ML/AI tools often carry the misleading aura of self-evident advanced technology and this falsely limits the perceived need for careful validation and verification of their performance, clinical and overall utility once they begin being used in the routine clinical practice. A recent systematic review demonstrated that only a couple of hundred randomized clinical trials (out of a total of almost a million) examined how the use of diagnostic tests affected clinical outcomes [78], and therefore demonstrated their clinical utility. From the ML/AI domain, the experience of Babylon Health's symptom checker-based triaging illustrates a similar point. In early testing with patient focus groups there were concern that there might be "[...] gaming [of] the symptom checker to achieve a GP appointment" [4]; an illustration of how algorithms are not always used as intended in the real world, and that these factors need to be assessed using pragmatic clinical trials [79].

This is further complicated by the fact that if ML/AI algorithms do not qualify as medical devices (as not all do), they are not subject to regulatory approval, and thus a large incentive to demonstrate their real world effectiveness is absent. The important consequence to be aware of in these situations, is the potential for patients [80], and the health systems in which the algorithm is utilised, being left exposed to unintended consequences. For example, a recently released heart age calculator appears to prompt all individuals over the age of 30, regardless of the absence of any additional risk factors, to visit their general practitioner to have their cholesterol and blood pressure checked [81]. The overwhelming concern is that this could result in a substantial increase in the number of visits to primary care physicians, in a situation where the evidence for the benefits this may reap have previously been found to be inadequate by the UK's healthcare guidance developing body [82].

## *Implementation-specific Questions*

### **17. How is the model being regularly re-assessed, and updated as data quality and clinical practice changes (i.e. post-license monitoring)?**

Even when there is sufficient evidence of efficacy and real-world effectiveness of a model to endorse its widespread use in clinical practice, it is important to recognise that the effectiveness requires constant review given the dynamic landscape of the healthcare environment. For example, computer-aided diagnosis programs have become an integral part of breast cancer screening programs around the world since the FDA first approved one for use in 1998 [83], but are they still as useful as they were 20 years ago? Most of the commercially available tools are based on neural networks which identify regions of interest, and diagnose the identified abnormality (e.g. calcification or mass). Early studies demonstrated modest increases in detection rates of breast cancer using CAD, compared to clinicians working without the aid of a CAD system [1,84]. However, almost 20 years following the FDA's first license for a mammography-based CAD system, national registry-based studies have produced sobering results, illustrating no significant improvement in diagnostic accuracy associated with CAD use in mammography interpretation [85]. Moreover, researchers have recently demonstrated that incorrect prompts from mammography-based CAD systems can actually decrease the sensitivity of more discerning users by up to 0.145 (95% CI, 0.034-0.257) for difficult cases [86]. Whilst the work discussed is not a comprehensive review of CAD in breast cancer, the results clearly suggest that there is a need to constantly re-evaluate technologies, as their usefulness can change over time. Other changes that might impact the healthcare landscape, include: changes in the case-mix of the population (e.g. demographic & eligibility changes), and improvements in treatments.

### **18. Is the ML/AI model cost-effective to build, implement, and maintain?**

While ML/AI algorithms may offer transformational benefits to healthcare systems and patients, there are significant costs associated with the development of software, the generation and use of data, implementation of a new system in practice, and acting on the additional information provided. Understanding the potential clinical benefit of new models - over and above current practice - alongside the cost (or savings) introduced by utilising these models should form part of any healthcare decision maker's appraisal of ML/AI-based technologies. Effective appraisal will require the development of assessment frameworks that take into account both the evidence for effectiveness and the evidence for economic impact. This is an area where healthcare decision makers, such as NICE, the FDA, etc., are crucial, as they can help developers of ML/AI models by providing clear guidance on the appropriate evidence that should be generated to demonstrate both effectiveness and economic impact, including: credible evidence relating to technical accuracy of the models, the relevant outcomes that demonstrate clinical effectiveness in general practice and, as appropriate, evidence to inform decision makers on the budget impact or the cost-effectiveness.

**19. How will the potential financial benefits be distributed if the ML/AI model is commercialized?**

Like all technologies ML/AI algorithms may have a market value. In situations where commercialisation is a goal, it is important to remember that health systems, and governmental research funding, can make a substantial contribution to the creation of an algorithm, via the associated sunk costs, such as data acquisition (clinicians' time, scanners, etc.), data annotation (training clinicians who eventually interpret the data generated), and sometimes even the developers' time (i.e. when they are publically funded researchers). In a publicly funded health system this even more important, as the symbiotic relationship between data-generating institutions and those with the capabilities to create AI/ML algorithms, is only possible because of expectation that benefits arising from the use of data will be retained (to some degree) by the health system, thereby satisfying the social contract with the public. Therefore, it is not only important to ensure that the investment and contributions of a health system and/or institution to an algorithm is recognised, but also that a mechanism be put in place to compensate them for having done so.

**20. How have the regulatory requirements for accreditation/approval been addressed?**

Software products including ML/AI algorithms can be subject to numerous regulatory requirements, depending on the setting in which the product will be used, from research and development to placing the product on the market (See box 2 for a high-level overview of the UK's regulatory framework). In our experience, whilst most clinicians are aware of CE marking of physical devices (the regulatory framework in the EU and the UK), its application to software products can often be a surprise. This is also true of software developers. Given that the regulatory landscape for health-related ML/AI-based software has changed dramatically over the last decade, and will continue to respond dynamically to innovation for the foreseeable future, it is important that discussions regarding the regulatory requirements for products in development are made early in the planning process. However, having this conversation once is clearly not sufficient, for example, devices that are developed and used in-house (in the UK) are not currently subject to device regulations, but this will change in 2020 when new updated regulations apply [27,28], and as such, regular review of regulatory compliance is necessary.

**Conclusion: from critical questions to a consensus TREE framework**

In much the same way that over previous decades' clinicians have been aided by frameworks for evaluating the strength of evidence, the ML/AI field should usefully build on what has been learned in traditional statistical approaches for 'clinical evidence' and the quality assurance pipeline (e.g. [6], [30], [90-94]). However, as we show here some of the challenges are new and different. Encouraging patients, clinicians, academics, and all manner of healthcare decision makers to ask the challenging questions raised by the TREE framework we propose here, will hopefully contribute to the development of safe and effective ML/AI-based tools in healthcare. By beginning to answer these questions in different settings, we can start to understand what constitutes a good answer, and we expect that the resulting discussion will be central to developing an international consensus framework for robust decision making in ML/AI at every stage of the research, translation, and application process in advanced 'health and wellbeing' data science. Developing such a framework will involve many challenges, including: finding common terminology (where key terms partly or fully overlap in meaning; balancing the need for robust empirical evidence of effectiveness without stifling innovation; identifying how best to manage the many open questions regarding best practices of development and communication of results; the role of different venues of communication and reporting; whilst simultaneously providing sufficiently detailed advice to produce actionable guidance for non-experts; and balancing the need for transparency against the risk of undermining IP rights. The path to a definitive solution will undoubtedly be long, and complex, but with the questions detailed above, we would argue that we have taken a step in the right direction. The next step will be to undertake a more comprehensive review of each of the 20 questions, to establish what is known with certainty, where the literature is lacking, and to identify high priority issues for the community to focus on. For those who would like to join us in this endeavour, we offer an open invitation to participate in the development of this AI-TREE framework.

# Box 2: A High-Level Overview of The UK's Regulatory Framework for ML/AI Algorithms in Health

Developers need to determine whether their ML/AI algorithm falls under the Medical Device Regulations' remit [87], which until 2010 did not regulate independent software products. These regulations cover products that make claims with a medical nature such as: providing diagnostic information; making recommendations for therapy; or, providing risk predictions of disease. The MHRA has published guidance for developers that covers this in greater detail [88]. If an algorithm does fall within the remit of the aforementioned regulation, the developer must then seek regulatory approval/accreditation in the form of a "CE" mark, prior to marketing it.

To CE mark an algorithm, the developer must follow one of the applicable conformity assessment routes, which, for the medium and high risk products, will require the involvement of a 'Notified Body' to assure the process. The developer must ensure that the device meets the relevant 'Essential Requirements' before applying the CE mark. These requirements include:

- Benefits to the patient shall outweigh any risks.
- Manufacture and design shall take account of the generally acknowledged state of the art.
- Devices shall achieve the performance intended by the manufacturer
- Software must be validated according to the state of the art taking into account the principles of development lifecycle, risk management, validation and verification
- Confirmation of conformity must be based on clinical data. Evaluation of this data must follow a defined and methodologically sound procedure.

In addition to the above, manufacturers are required to have post market surveillance provision to review experience gained from device use and to apply any necessary corrective actions.

Moreover, the use of ML/AI algorithms may be regulated indirectly by other legislation or regulatory agencies. Perhaps the highest profile additional legislative framework to be aware of is the EU General Data Protection Regulations (GDPR), the relevance of which has been discussed elsewhere (questions 3 and 16). In terms of other regulatory agencies who play an important role in the regulation of ML/AI software in health, the UK's Care Quality Commission (CQC) is one group to be aware of, as they are tasked with monitoring compliance with NHS Digital's Clinical risk management standards [89]; a contractual requirement placed on developers engaging in service provision to the NHS.

**Acknowledgements**

The authors would like to thank all those at The Alan Turing Institute, HDR UK, the National Institute for Clinical and Care Excellence (NICE), the Medicines and Healthcare products Regulatory Agency (MHRA), the Clinical Practice Research Datalink (CPRD), the Enhancing the Quality and Transparency of Health Research (EQUATOR) Network, the Meta-Research Innovation Centre at Stanford (METRICS), and the Data Science for Social Good (DSSG) Program at the University of Chicago, who supported this project.

**Funding**

SJV, BAM, GB, FJK, & CH are employees of The Alan Turing Institute supported by the EPSRC grant EP/N510129/1. GSC was supported by the NIHR Biomedical Research Centre, Oxford. HH is a National Institute for Health Research (NIHR) Senior Investigator. His work is supported by: 1. Health Data Research UK (grant No. LOND1), which is funded by the UK Medical Research Council, Engineering and Physical Sciences Research Council, Economic and Social Research Council, Department of Health and Social Care (England), Chief Scientist Office of the Scottish Government Health and Social Care Directorates, Health and Social Care Research and Development Division (Welsh Government), Public Health Agency (Northern Ireland), British Heart Foundation and Wellcome Trust. 2. The BigData@Heart Consortium, funded by the Innovative Medicines Initiative-2 Joint Undertaking under grant agreement No. 116074. This Joint Undertaking receives support from the European Union's Horizon 2020 research and innovation programme and EFPIA; it is chaired, by DE Grobbee and SD Anker, partnering with 20 academic and industry partners and ESC. 3. The National Institute for Health Research University College London Hospitals Biomedical Research Centre. Dr Ioannidis is supported by an unrestricted gift from Sue and Bob O'Donnell to the Stanford Prevention Research Center. The Meta-Research Innovation Center at Stanford (METRICS) is supported by a grant from the Laura and John Arnold Foundation. KGMM is supported by The Netherlands Organisation for Heath Research and Development. GB and SJV are supported by the University of Warwick's EPSRC-funded Impact Acceleration Account (IAA). PJ, SC, KSLM, & AJ are employees of the National Institute for Clinical and Care Excellence (NICE). PM, DG, MB, & RB are employees of the Medicines and Healthcare products Regulatory Agency. SJV is supported by the data study group funding as its director (/TU/B/000012).